\documentclass[aps,prb,superscriptaddress,preprintnumbers,twoside]{revtex4}
\usepackage{bm}
\usepackage{graphicx}
\usepackage{amsfonts}
\usepackage{amsmath}
\usepackage{amssymb}
\newcommand {\be}{\begin{equation}}
\newcommand {\ee}{\end{equation}}
\newcommand {\bea}{\begin{eqnarray}}
\newcommand {\eea}{\end{eqnarray}}

\begin{document}

\title{Optimization of the design of superconducting inhomogeneous
nanowires}
\author{Ilya Grigorenko}
\affiliation{Theoretical Division T-11, Center for Nonlinear
Studies, Center for Integrated Nanotechnologies, Los Alamos
National Laboratory, Los Alamos, New Mexico 87545, USA}
\author{Jian-Xin Zhu}
\affiliation{Theoretical Division T-11, Los Alamos National
Laboratory, Los Alamos, New Mexico 87545, USA}
\author{Alexander Balatsky}
\affiliation{Theoretical Division T-11, Center for Nonlinear
Studies, Center for Integrated Nanotechnologies, Los Alamos
National Laboratory, Los Alamos, New Mexico 87545, USA}
\date{\today}

\begin{abstract}
We study optimization of superconducting properties of
inhomogeneous nanowires. The main goal of this research is to find
an optimized geometry that allows one to maximize the desired
property of superconductors, such as the maximum value of local
superconducting gap or total condensation energy. We consider
axially symmetric design of multi-layered nanowires with
possibility to adjust and change the layers thickness. We use
numerical solution of the Bogoliubov-de Gennes equations to obtain
the local superconducting gap for different arrangements of the
inhomogeneous structures.  The value of the optimized properties
can be up to $300\%$ greater compared to a non-optimized geometry.
The optimized configuration of multilayers strongly depends on the
desired property one wants to optimize and on the number of layers
in the nanowire.
%Possible experimental
%realizations of the proposed structures are discussed.
\end{abstract}
\maketitle
\section{Introduction}

Recent advances in fabrication and experimental measurement of
spatially inhomogeneous superconducting materials open up new
opportunities in optimal design of targeted properties of
superconducting materials. Optimal design is an approach where one
calculates the properties of a given structure or device and then
optimizes the shape, composition or mutual orientation of the
parts of the structure with the purpose of achieving some
function, some target property of the structure. With proper
chosen geometry (that determines boundary conditions for quantum
systems) one can expect enhancement of the target property at
certain location, or even observe a ''quantum mirage" in confined
corral-like nanoscale systems \cite{coral}.

The ideas of optimal design are following the same line of
thinking as the band structure engineering \cite{zunder1,zunder2},
or prediction of the complex materials structure using
information-theoretical methods \cite{ceder}. Notion of optimal
design also can be viewed as an extension of similar ideas to the
design of decoherence-protected quantum state engineering in
quantum computing \cite{eberly,ilya}.

%Superconductivity is a macroscopic quantum phenomena.
In this work we propose to extend the notion of design of quantum
states to the case of superconducting materials. We will focus on
the superconducting nanostructures. Examples of the specific
parameters that one might want to optimize in superconducting
materials are local superconducting gap $\Delta({\bf r})$ and
maximum of the condensate energy $E_{cond}\propto \int d{\bf r}
|\Delta({\bf r})|^2$. These are the main target parameters we will
focus on below. Specifically, we are going to investigate the
interplay between strong spatial inhomogeneities, confinement and
geometry effects in superconducting nanoscaled systems. To
elucidate this interplay we will focus on the influence of the
spatial arrangement of layers of different superconducting
materials on the target parameters we mentioned above.

Obviously, the search for optimal design can only be performed
using physically correct and computationally accessible model. In
our approach we use the Bogoliubov-de Gennes equations which is
well-suited to describe inhomogeneous superconductors
microscopically. Since three-dimensional spatial inhomogeneity is
not tractable without using a supercomputer, we resort on the case
of cylindrical quasi-one-dimensional systems, considering axially
symmetric nanowires. The axially symmetric superconducting wires
were recently fabricated and have shown their usefulness for
prospective measuring devices
\cite{nanowire,nanowire1,nanowire2,nanowire3}.

Before we proceed with the results we will make a few technical
comments. Spatial inhomogeneity destroys translational symmetry
and can create localized states on a certain length scale. If the
inhomogeneities are created by disorder, it usually may be
characterized by simple, low order, and spatially localized
correlation functions. However, this approach cannot be applied in
the case of engineered, ordered inhomogeneities. In this case
inhomogeneities are characterized by spatially delocalized
correlation functions of high order. As a result, this problem is
not tractable analytically. One needs to use numerical solutions
with a fine mesh and at the same time allow for variations in the
solution as we search for optimized configuration. This approach
was not possible computationally ten years ago and only now
becomes feasible.

In our approach we abandon the usual assumption that the designed
inhomogeneity can be treated as a small perturbation. Detailed
analysis can demonstrate that the size of the space of accessible
solutions grows with the maximum allowed amplitude of the
inhomogeneities. Thus, a relatively big inhomogeneity may bring
new effects and functionalities, which are not accessible by small
perturbations. The assumption of relatively large spatial
inhomogeneities place our search for optimal configurations beyond
the realm of applicability of the Anderson's theorem. The
Anderson's theorem for non-magnetic alloys and zero external field
states that to a first approximation the excitation spectra are
the same for alloy as for the pure metal \cite{anderson}. This
theorem resorts on the assumptions of small spatial inhomogeneity
of the pair potential $\Delta(r)$ and chemical similarity of the
introduced impurities. Both of the assumptions do not hold in
configurations that turned out to be optimal.

If designed inhomogeneities are large, a naive strategy to
maximize the local superconducting gap $\Delta({\bf r})$ may be
achieved by controlling quasiparticle local density of states at
and near the Fermi level. On a deeper level one may expect the
interplay between correlations of quasiparticles on the scale of
the coherence length $\zeta$ and the scale of spatial
inhomogeneity. Note that the coherence length may be also modified
by the spatial inhomogeneity, thus the problem of the finding of
the proper length scales should be solved self-consistently.

Important effect that comes into play is the confinement effect
\cite{cyl1}, when a quasiparticle subband appearing due to the
finite size quantization happens to be close to the Fermi surface.
In this case the density of states is increased, that leads to
enhancement of the pairing potential.

The pairing potential $\Delta({\bf r})$ may also depend of the
finite size of the system via the phonon quantization. However, as
it was shown for thin films \cite{sarma}, this effect is a minor
correction to the quasiparticle quantization mechanism mentioned
above. Here we leave consideration of the phonon quantization
effects for future publications and consider phonons to be the
same as in the bulk material.

This paper is organized as follows. In Section II we present an
approach, based on numerical solutions of the Bogoliubov-de Gennes
equations. In Section III we discuss the results of numerical
solutions and present main results. In Section IV we discuss main
qualitative finding of the optimal design approach and its
applicability to other problems. We conclude in Section V with the
discussion of main results and the outlook of future approaches.

\section{Approach and basic facts}

In this work we use Bogoliubov-de Genes formalism \cite{bdg_eq}.
It is a mean-field formulation of microscopic theory of weak
coupling superconductors, and it is particularly well suited for
spatially inhomogeneous problems. It is also appropriate for the
relatively small heterostructures in the clean limit that we are
interested in.

Using the microscopic theory for inhomogeneous superconductor, one
can explore the rich structure in the local density of states
(LDOS), which now can be measured directly by using STM
\cite{STM}. Such STM studies, unlike the previous experimental
studies \cite{before_STM} which explored only spatially averaged
properties, may provide valuable information about local
variations of the pair potential $\Delta({\bf r})$.

We are going to use real space discretization technique
\cite{salomaa}, that unlike using basis of Bessel functions for
the cylindrically symmetric problems \cite{cyl1,maki}, leads to
the eigenproblem with sparse matrices. In this case the
eigenproblem solution can be found much faster using sparse
iterative solvers. In this work we choose the wire to be hollow to
simplify numerical calculations and to avoid uneven real space
descretization for the numerical treatment of the singularity near
the origin. However, our results will not be changed qualitatively
in the case of solid wires of the same size.

The Bogoliubov - de Gennes (BdG) equations for the quasiparticle
amplitudes $u_n$ and $v_n$ with excitation energy $E_n>0$ can be
written in a compact form \bea \label{BDG} ( -\frac{\hbar^2}{2
m^*}\nabla^2-\mu )\hat{\sigma}_z \hat{\psi}_n({\bf
{r}})+\Delta({\bf {r}})\hat{\sigma}_x \hat{\psi}_n({\bf {r}})=E_n
\hat{\psi}_n({\bf {r}}), \eea where we denote $\hat{\psi}_n\equiv
(u_n,v_n)^T$, and $\hat{\sigma}_x,\hat{\sigma}_z$ are the Pauli
matrices. Here $\mu$ is the chemical potential. Any effects of the
band structure of the material are included in the quasiparticle
effective mass $m^*$. In the presence of magnetic field one should
replace the Laplace operator by $({\bf p}- \frac{e}{c}{\bf A})^2$.

The superconducting pair potential $\Delta({\bf r})$ for {\it
s}-type superconductor must be determined self-consistently from
the solutions of the BdG equations, as: \bea
\label{self_consistent} \Delta({\bf{r}})=g({\bf r})\sum_n u_n({\bf
r})v_n^*({\bf r})[1-2f(E_n)] \Theta( \hbar \omega_D-E_n), \eea
where $f(E_n)$ is the Fermi distribution function, $\omega_D$ is
the Debye cut-off frequency, $\Theta()$ is the step function, and
$g({\bf r})$ is the inhomogeneous coupling constant. The summation
is performed over the indices $n$ corresponding to $E_n>0$.
Throughout this work, we use chemical potential $\mu=0$ and
temperature $T=0$K. The coupling constant is chosen in order to
obtain the bulk value $\Delta_{bulk}\approx 10$ K. Note that as in
other studies \cite{zhu}, similar model parameters were chosen for
illustrative purposes only, and were not intended to reproduce
realistic materials.

%The concentration of the quasipaticles is determined by
% \bea \label{chemical}
% n=\frac{2}{V}\int dV \sum_n[|u_n({\bf r})|^2 f(E_n)+|v_n({\bf
% r})|^2(1-f(E_n))],
% \eea
% where $V$ is the volume of the system.
Assuming $z$ axis to be an axis of the axial symmetry in the wire
of length $L$ and imposing periodic boundary conditions along this
axis, the solution of Eq. (\ref{BDG}) has the form: \bea
\label{BDG_solution}
\hat{\psi}_n(\rho,\phi,z)=\hat{a}_n(\rho)\exp(i
l\phi)/\sqrt{2\pi}\exp(i k_z z)/\sqrt{L}. \eea The BdG equations
for the radial part can be written as: \bea \label{BDG_cyl} (
-\frac{\hbar^2}{2 m^*}\nabla_\rho^2-\mu )\hat{\sigma}_z
\hat{a}_n(\rho)+\Delta(\rho)\hat{\sigma}_x \hat{a}_n(\rho)=E_n
\hat{a}_n(\rho), \eea where we use the notation \bea
\label{radial} \nabla_\rho^2\equiv \partial^2/\partial \rho^2+
\rho^{-1}\partial/\partial \rho -l^2\rho^{-2}-k_z^2 \eea for the
radial part of the Laplace operator. Indices $l$ and $k_z=2\pi
n_z/L$ are the azimuthal quantum number and the wave vector in the
$z$ direction correspondingly.

To compute the self-consistent solution to both
Eqs.(\ref{BDG},\ref{self_consistent}), we assume an initial guess
$\Delta_0(\rho)=\Delta_{bulk}$ for the order parameter. Then by
means of numerical diagonalization, we compute the solutions
$u_i(\rho),v_i(\rho)$, for each combination $\{l,k_z\}$,
corresponding to $\Delta_0(\rho)$. We use these solutions to
generate new guess for the order parameter.

In order to avoid instabilities during the self-consistent
calculations, that we found typical for strongly inhomogeneous
geometries, we use a simple mixing procedure: \bea \label{mixing}
\Delta_{mix}(\rho)=\alpha\Delta_{i}(\rho)+(1-\alpha)\Delta_{i-1}(\rho),
\eea where $\alpha$ is an adjustable parameter, $0<\alpha\le 1$.
Initially we set $\alpha=0.5$.
 To insure convergence, we increase the current value of
  $\alpha$ by $5\%$ if the relative deviation between two consequent steps
   $S^i=\max_{\rho}|\Delta_i(\rho)- \Delta_{i-1}(\rho)|/
   \max_{\rho}|\Delta_i(\rho)|$ has
   decreased,  $S^i<S^{i-1}$. And we decrease $\alpha$ by $20\%$ in the opposite
   case, $S^i>S^{i-1}$.\\
The computed $\Delta_{mix}(\rho)$ is used for the next iteration
step.

 We repeat iterations until we achieve the
acceptable level of accuracy ($S^i<10^{-3}$). After the end of the
procedure, we perform an additional step with $\alpha=1$ to ensure
convergence of the  obtained solution. It usually takes $50-100$
iterations to converge.
%The coherence length
%\bea \label{coherence}
%\ksi_0=\hbar v_F/\Delta_{bulk}
%\eea
%p_F\ksi_0~10^3$

\section{Results}

\subsection{Homogeneous wires} In order to check our simulations we
first reproduce the bulk limit by considering a relatively big
system. We assume a hollow homogeneous cylindrical wire with zero
value boundary conditions on the inner and outer surfaces. For
such small systems the role of the boundary conditions, as we see
later, is crucial. Zero boundary conditions we chose does not
imply necessarily contact with a normal metal. It can be  vacuum
(or just air) outside the wire.

\begin{figure}[h] %FIGURE 2
%\vspace{1.cm}
\includegraphics[width=10.cm, angle=0,bb = 1 1 670 700]{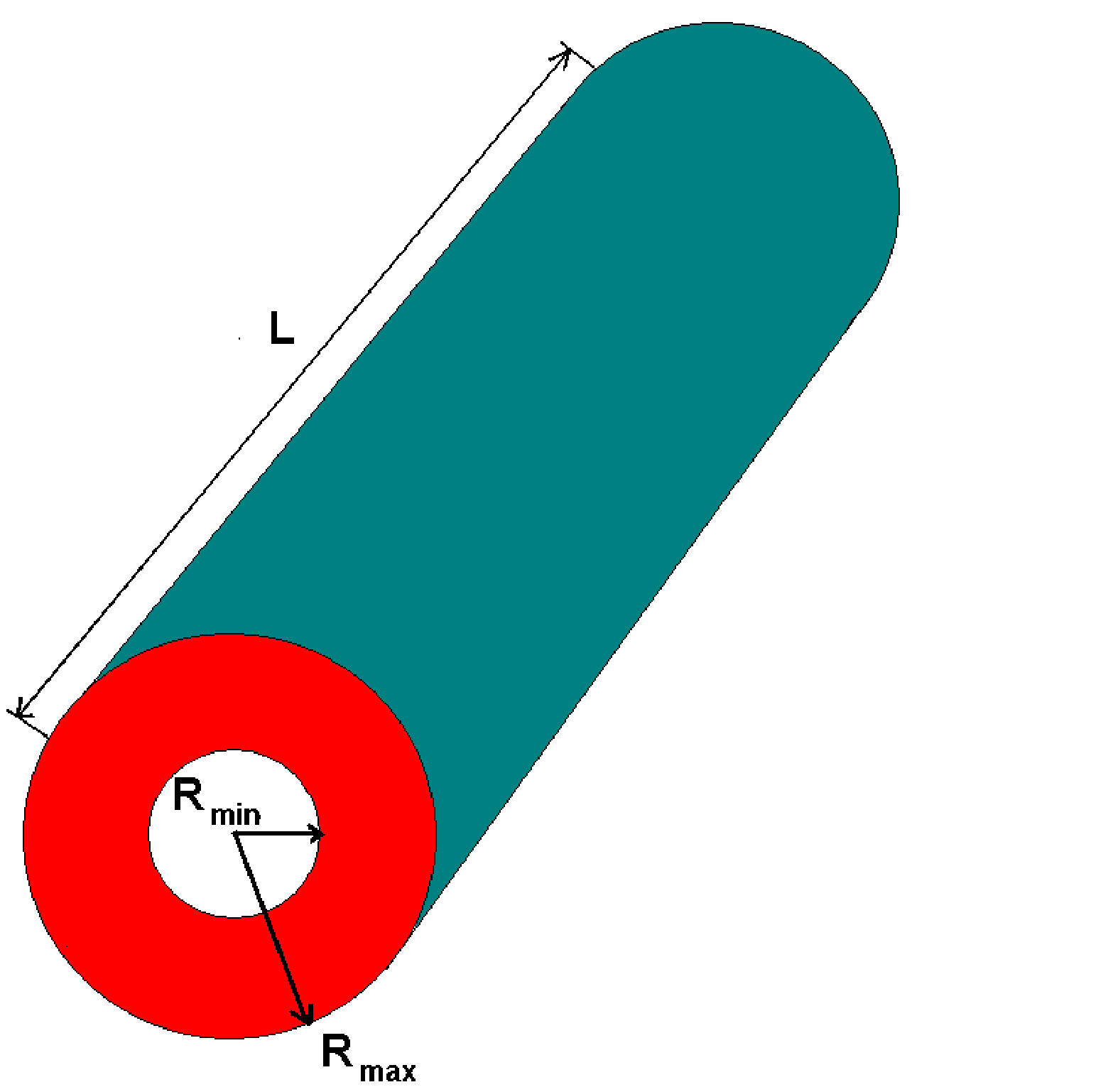}
\includegraphics[width=10.cm, angle=-90]{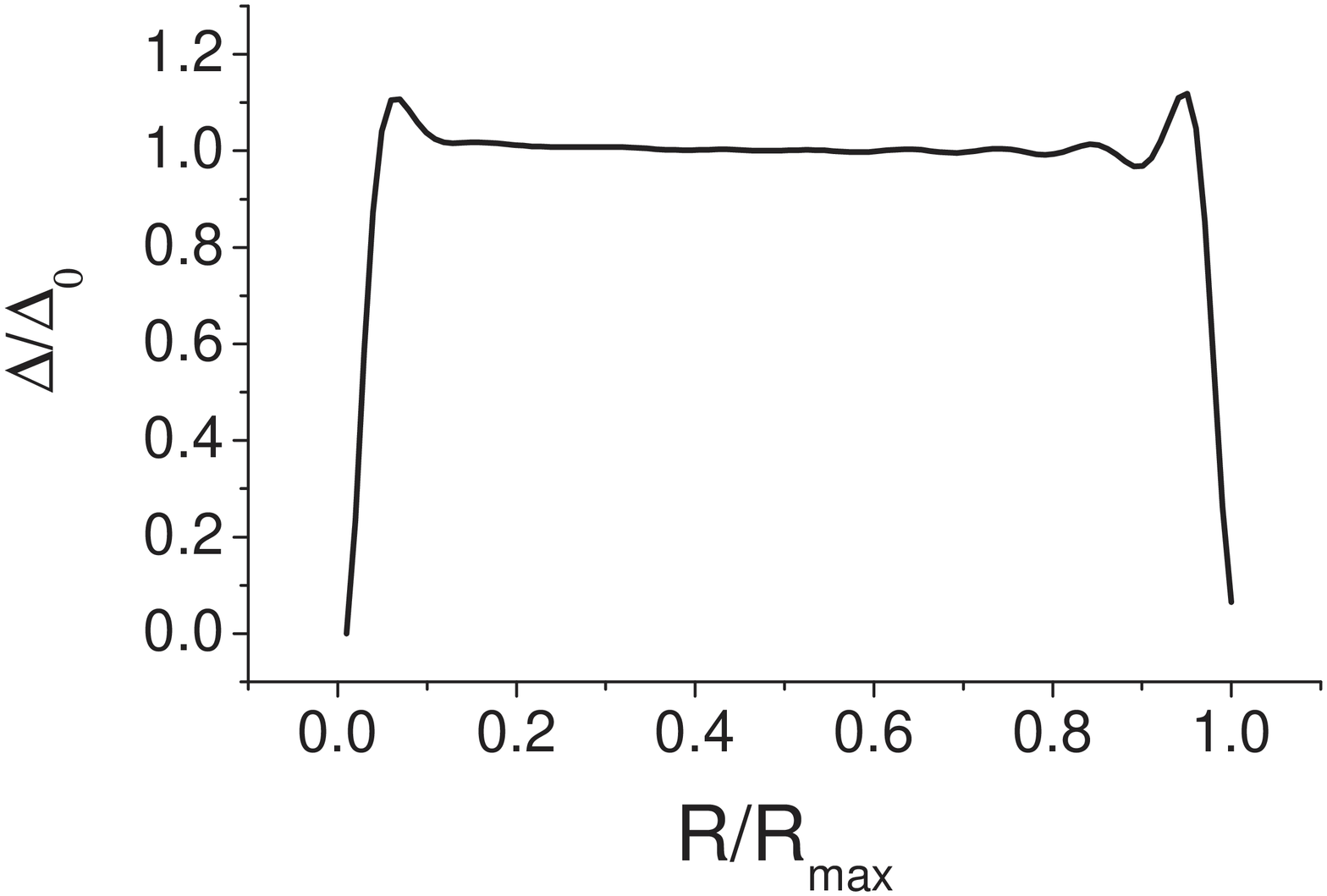}
\caption{\label{fig8} \footnotesize{The geometry of the system is
depicted in Fig. 1a and is represented by an axially symmetric
hollow wire. The calculated radial part of the superconducting
order parameter $\Delta(\rho)$ (see Fig. 1b), is normalized to the
bulk value, for a uniform infinite hollow superconducting wire
with inner radius $R_{min}=0.5$ nm and outer radius $R_{max}=50$
nm. The length of the wire is chosen to be $500$ nm. For our
calculations we assume temperature $T=0$ K. }}
\end{figure}

 We choose the outer radius of the wire
$R_{max}=50$ nm, and the inner radius as $R_{min}=0.5$ nm (see
Fig. \ref{fig8}). The length of the wire is set to be $L=500$ nm
with periodic boundary conditions that is a sufficient
approximation for an infinitely long wire. The temperature in all
our simulations is set to zero $T=0$ K.

%For this geometry we would like to exclude volume-related
%effects. For this purpose we consider the central layer with a
%variable thickness, that results in the constant volume.
%we focus on few geometries...describe heer the geoemetries first.
%move here any discussion that is main subject. keep technical
%discussion to previous general set up chapter.

\begin{figure}[h] %FIGURE 2
%\vspace{2.cm}
\includegraphics[width=9.cm, angle=0,bb =  200 400  1300 1300]{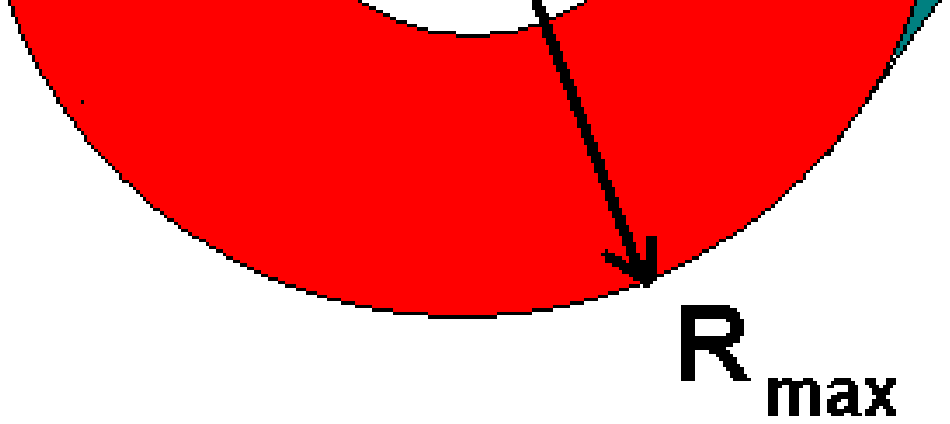}
%\vspace{-2.cm}
\includegraphics[width=10.cm, angle=0]{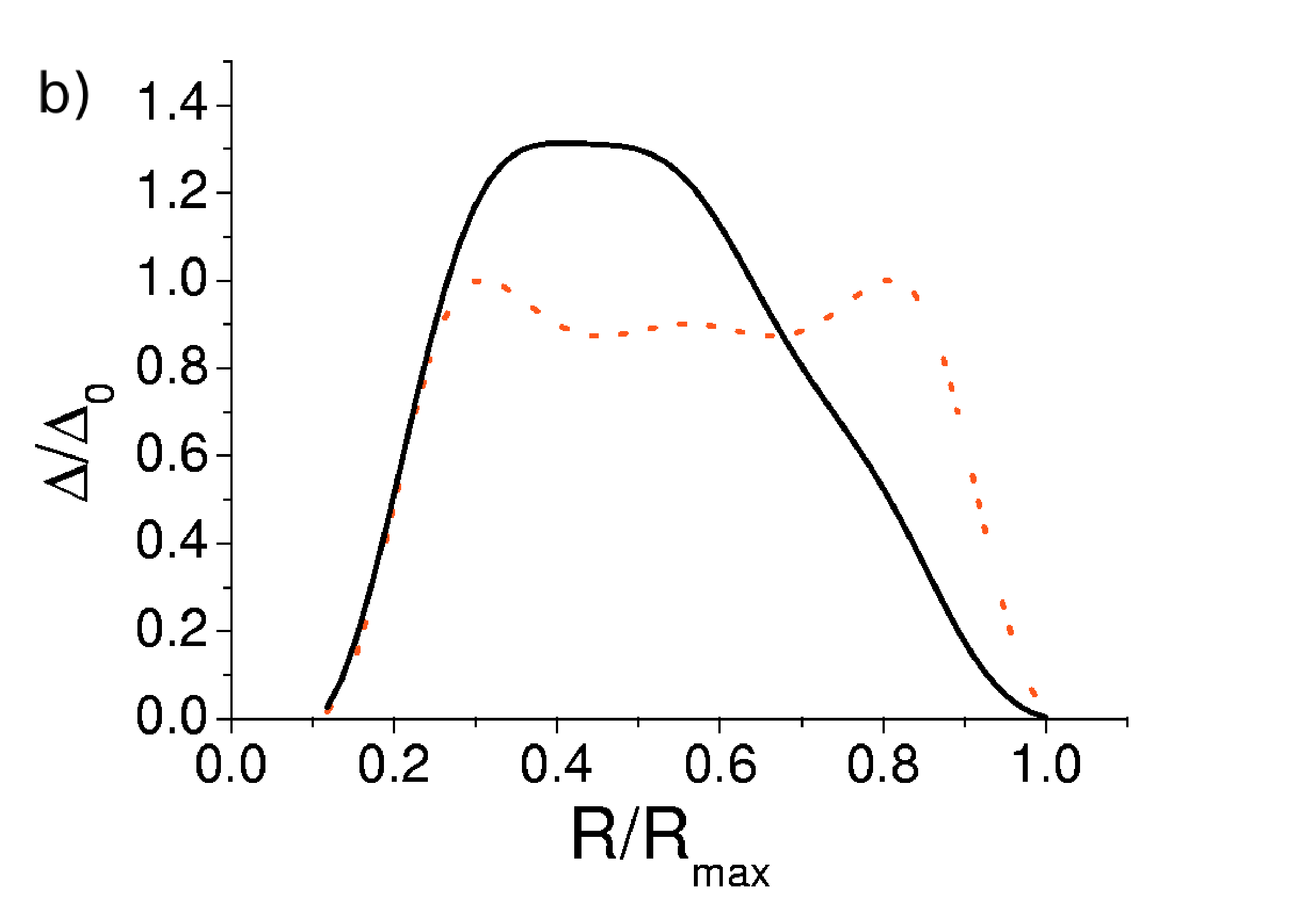}
%\vspace{1.cm}
\caption{\label{fig1} \footnotesize{The geometries
of the systems under consideration are depicted in Fig. 2a and are
represented by a hollow wire and a thin film. The calculated
radial part of the superconducting order parameter $\Delta(\rho)$
(see Fig. 2b), for a uniform hollow superconducting wire with
inner radius $R_{min}=1.5$ nm and outer radius $R_{max}=15$ nm
(solid line). The length of the wire is chosen to be $500$ nm, and
the superconducting order parameter for uniform superconducting
film with the thickness $D=13.5$ nm (dashed line). The two other
dimensions of the film are set to $500$ nm. For our calculations
we assume temperature $T=0$ K. }}
\end{figure}

\begin{figure}[h] %FIGURE 2
%\vspace{1.cm}
\includegraphics[width=7.cm, angle=-90]{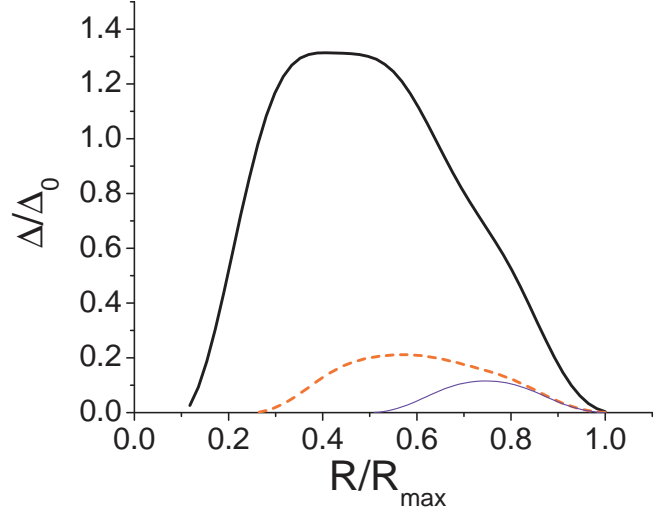}
\caption{\label{fig2} \footnotesize{The calculated radial part of
the superconducting order parameter $\Delta(\rho)$, for uniform
infinite hollow superconducting wire with the inner radii
$R_{min}$=$1.5$, $3.75$, $7.5$ nm and the constant outer radius
$R_{max}=15$ nm. The length of the wire is chosen to be $500$ nm
Note decrease of the order parameter with the decreasing of the
thickness. For our calculations we assume temperature $T=0$ K. }}
\end{figure}

We find that the radial part of the self-consisted pairing
potential $\Delta(\rho)$ is close to the bulk value with some
deviations near the boundaries. The slight rise of the pairing
potential can be attributed to the local approximation for the
pairing potential in the BdG equations, so it has a nature of the
Gibb's phenomenon. Now let us compare a hollow homogeneous
cylindrical wire with inner radius $R_{min}$ and outer radius
$R_{max}$, assuming zero boundary conditions, and a homogeneous
film of thickness $D=R_{max}-R_{min}$. The results of the
calculations are shown in Fig. \ref{fig1}. The spatially dependent
superconducting order parameter is normalized to the maximum value
$\Delta_0$ for the slab. We chose $R_{max}$=15 nm and
$R_{min}$=1.5 nm. For all our simulations we use the length of the
wire $L=500$ nm with periodic boundary conditions. The maximum
value of the $\Delta(\rho)$ for the hollow wire is about $30\%$
larger than for a slab. Clearly, for the wire the superconducting
order parameter has asymmetric shape due to different local
curvature at different values of $\rho$. Next we calculate
$\Delta(\rho)$ for wires of different thickness. In Fig.\ref{fig2}
we plot the superconducting order parameter $\Delta(\rho)$ for the
fixed outer radius $R_{max}$=15 nm and variable inner radius
$R_{min}$=1.5,\;3.75,\;7.5 nm. We found that
$\max\{\Delta(\rho)\}$ monotonously decreases as a function of
$R_{min}$, together with decreasing of the local curvature.

\subsection{Inhomogeneous wires}

In this section we focus on few geometries of inhomogeneous
nanowires. In the first geometry we consider a hollow axially
symmetric nanowire composed of two layers of two different
superconducting materials. We model different materials by using
different coupling constant $g$. We have chosen the thicknesses of
the layers to be equal, i.e. the first arrangement is
$g(\rho)=g_1$, for $R_{min}<R<(R_{min}+R_{max})/2$, and
$g(\rho)=g_2$, for $(R_{min}+R_{max})/2<R<R_{max}$.  The second
arrangement one can obtain by switching the constants $g_1$ and
$g_2$.

%If we assume that $g_1>g_2$, than an intuitive guess will be that
%the maximal value of  $\Delta(r)$ is larger for the second
%arrangement, since in this the volume of a better superconductor
%is bigger. However, we are going to show that the geometry effects
%lead to a counterintuitive result.

%The second geometry we consider is a three-layer geometry, namely
%$g(\rho)=g_1$, $R_{min}$ $g(\rho)=g_3$, $R_2$

%Now let us consider inhomogeneous superconducting nanostructures.
%We investigate two arrangements of the inhomogeneity.

\begin{figure}[h] %FIGURE 2
%\vspace{1.cm}
\includegraphics[width=7.cm, angle=0]{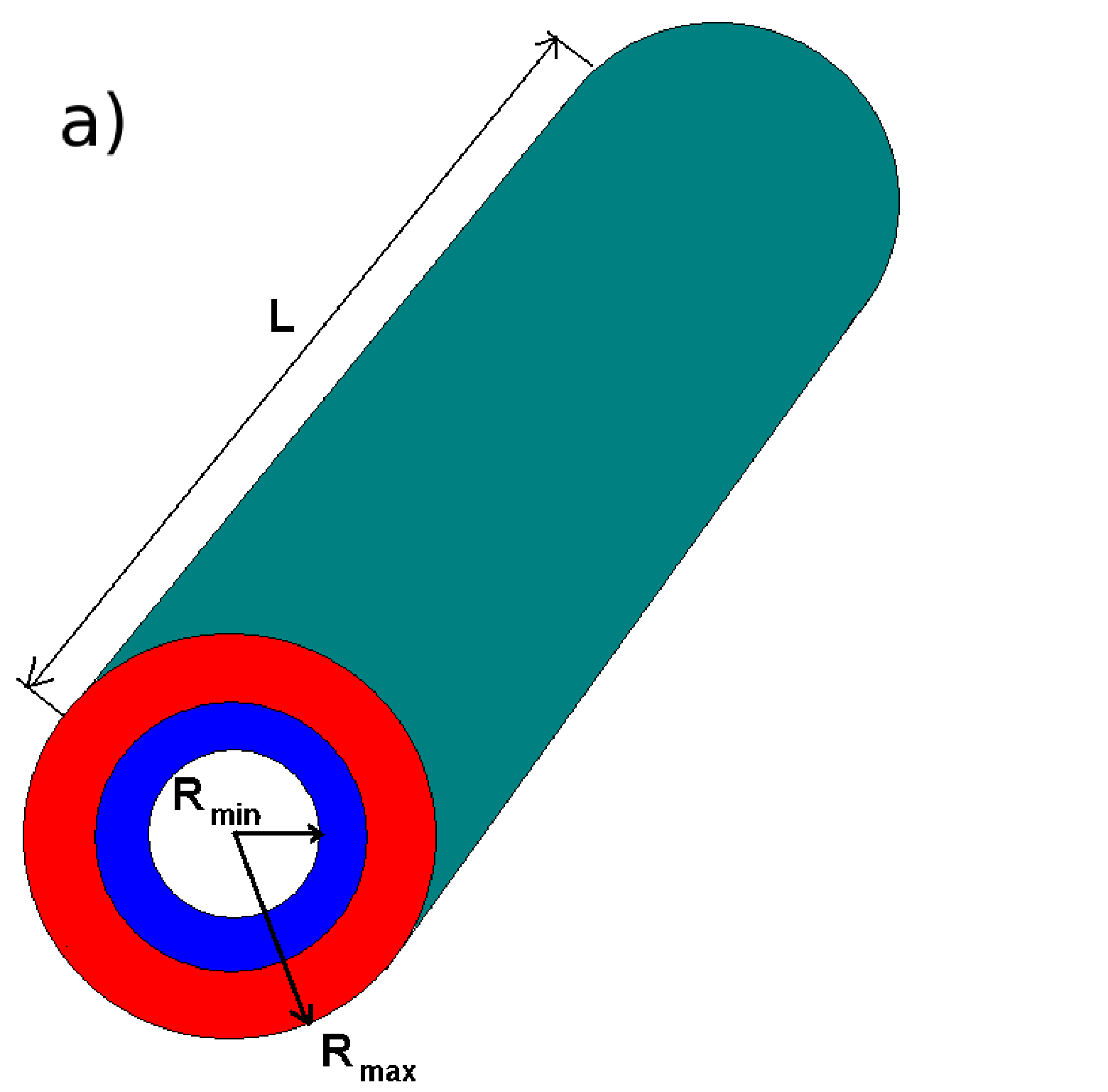}
\includegraphics[width=10.cm, angle=0]{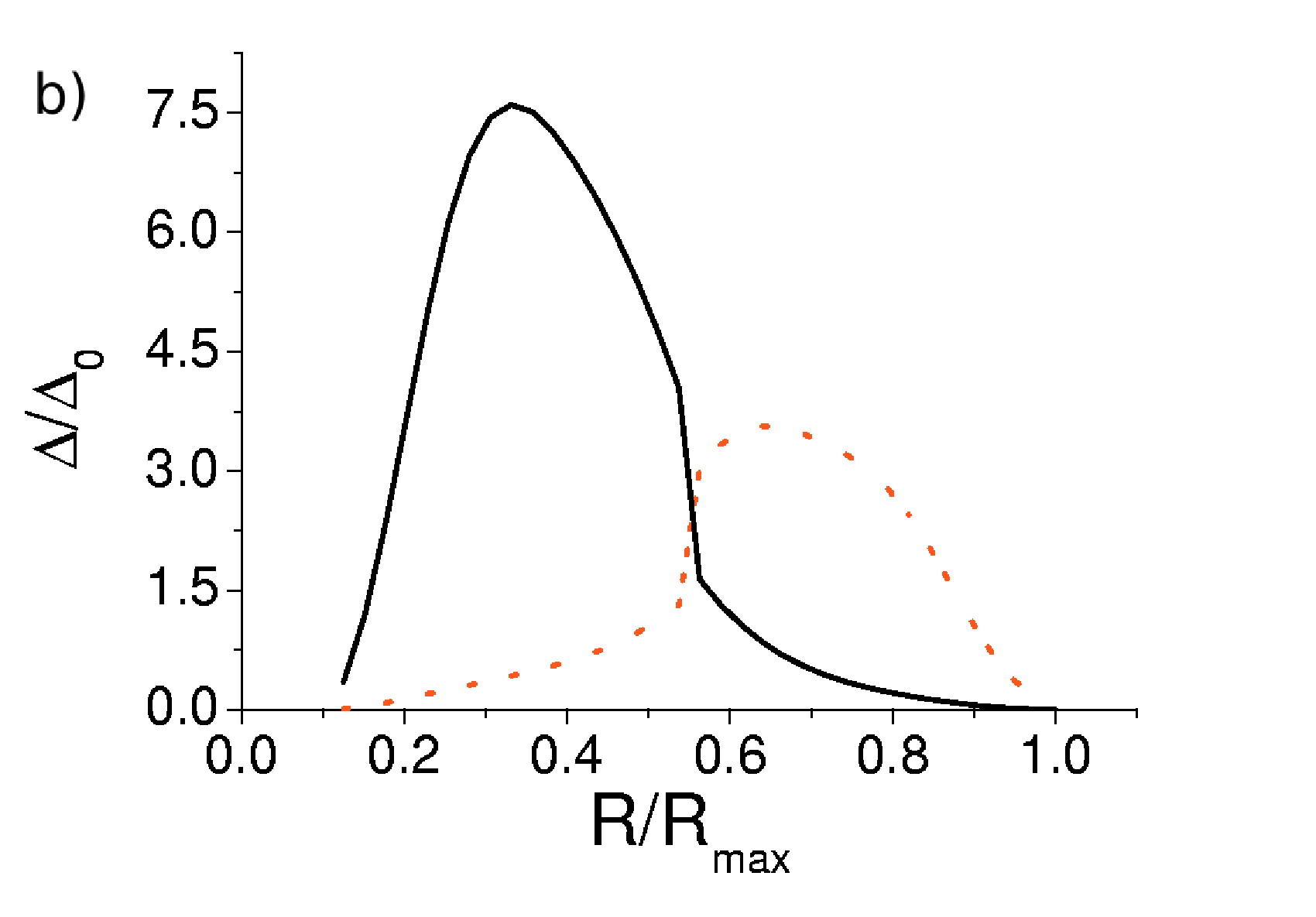}
\caption{\label{fig3} \footnotesize{The geometry of the system is
depicted in Fig. 3a and consists of $2$ axially symmetric layers.
The order parameter $\Delta(\rho)$(see Fig. 3b) for a nonuniform
hollow wire with $R_{min}=1.5 nm$ and $R_{max}=15$ nm. The length
of the wire is chosen to be $500$ nm. For our calculations we
assume temperature $T=0$ K. The thick solid line corresponds to
the case when the layer with stronger coupling constant $2g$
closer to the center, dashed line corresponds to the case when the
layer with stronger coupling constant close to the outer surface.
The weaker superconductor has the coupling constant $g$. Each
layer thickness $6.75$ nm. Note the non equivalence of the
arrangements due to the local curvature. }}
\end{figure}

Now let us demonstrate an unexpected effect of the geometry on the
superconducting order parameter of an inhomogeneous system. We
consider a wire which has an inner coaxial layer of a better
superconductor with the coupling constant $g_1=2g$ and outer layer
with the coupling constant $g_2=g$. For the second arrangement we
consider the inverse sequence of the layers, namely, the wire has
inner coaxial layer of superconductor material with the coupling
constant $g_1=g$ and outer layer with the coupling constant
$g_2=2g$. The first choice is depicted in Fig.\ref{fig3} by solid
line, and second choice - by dashed line. Note that the calculated
$\Delta(\rho)$ is normalized to maximum value $\Delta_0$ for the
slab geometry of the same thickness $D=R_{max}-R_{min}$. Note,
that because of the spacial variation of the local curvature the
exchange of the superconducting layers leads to nonequivalent
results. Interestingly, although for the second arrangement the
system has a bigger volume of the better superconductor, than in
the first arrangement, the maximum value of the pair potential
$\Delta(\rho)$ in this case is smaller. This effect is due to
local enhancement of the quasiparticles coupling in the effective
potential $U_{eff}\propto \rho^{-2}$ for $\rho\to 0$, originated
from the angular momentum term in the radial part of the Laplacian
Eq.(\ref{radial}).

\begin{figure}[h] %FIGURE 2
\includegraphics[width=5.cm, angle=0, bb= 84 220 437 625]{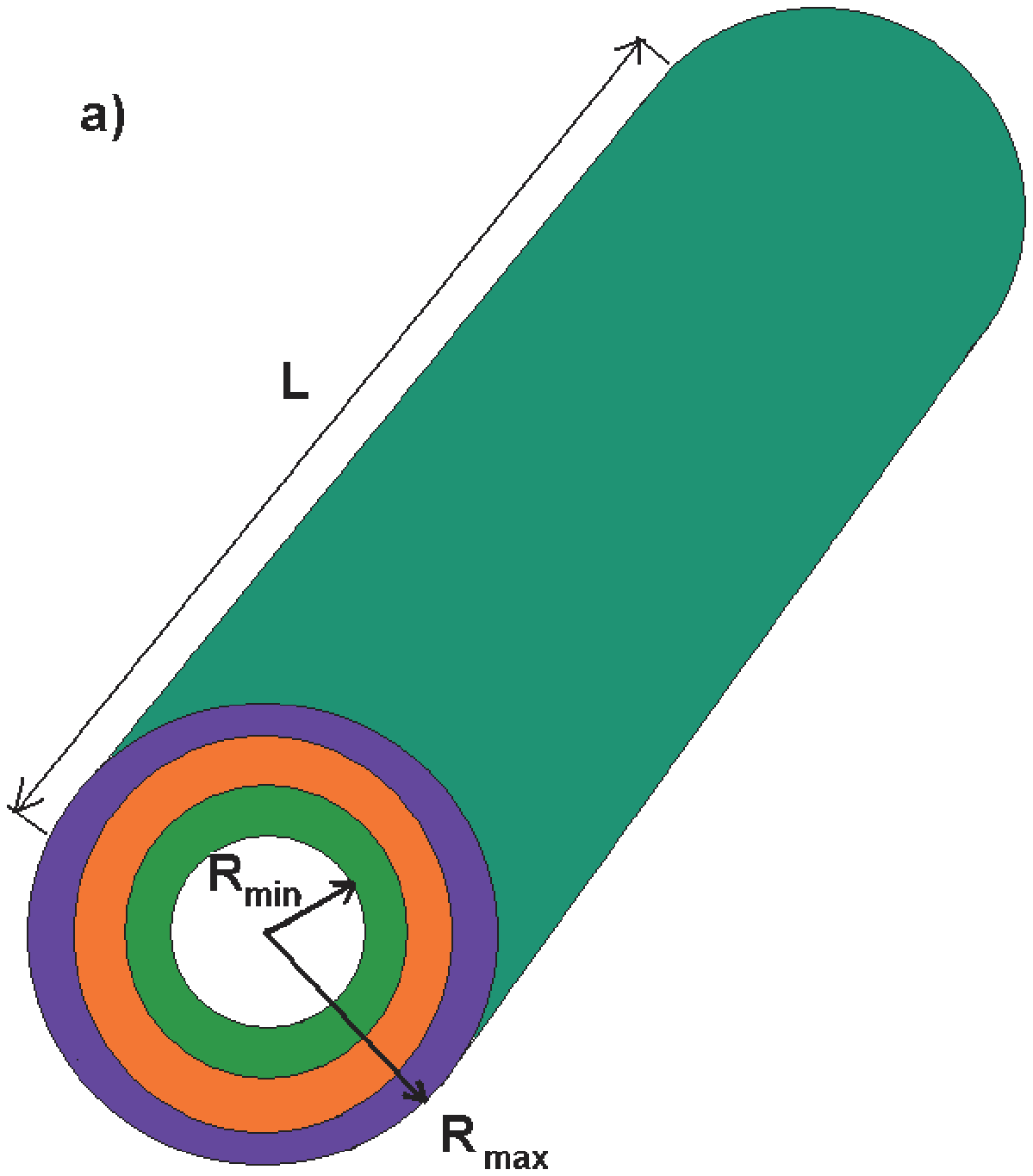}
\includegraphics[width=8.cm, angle=0]{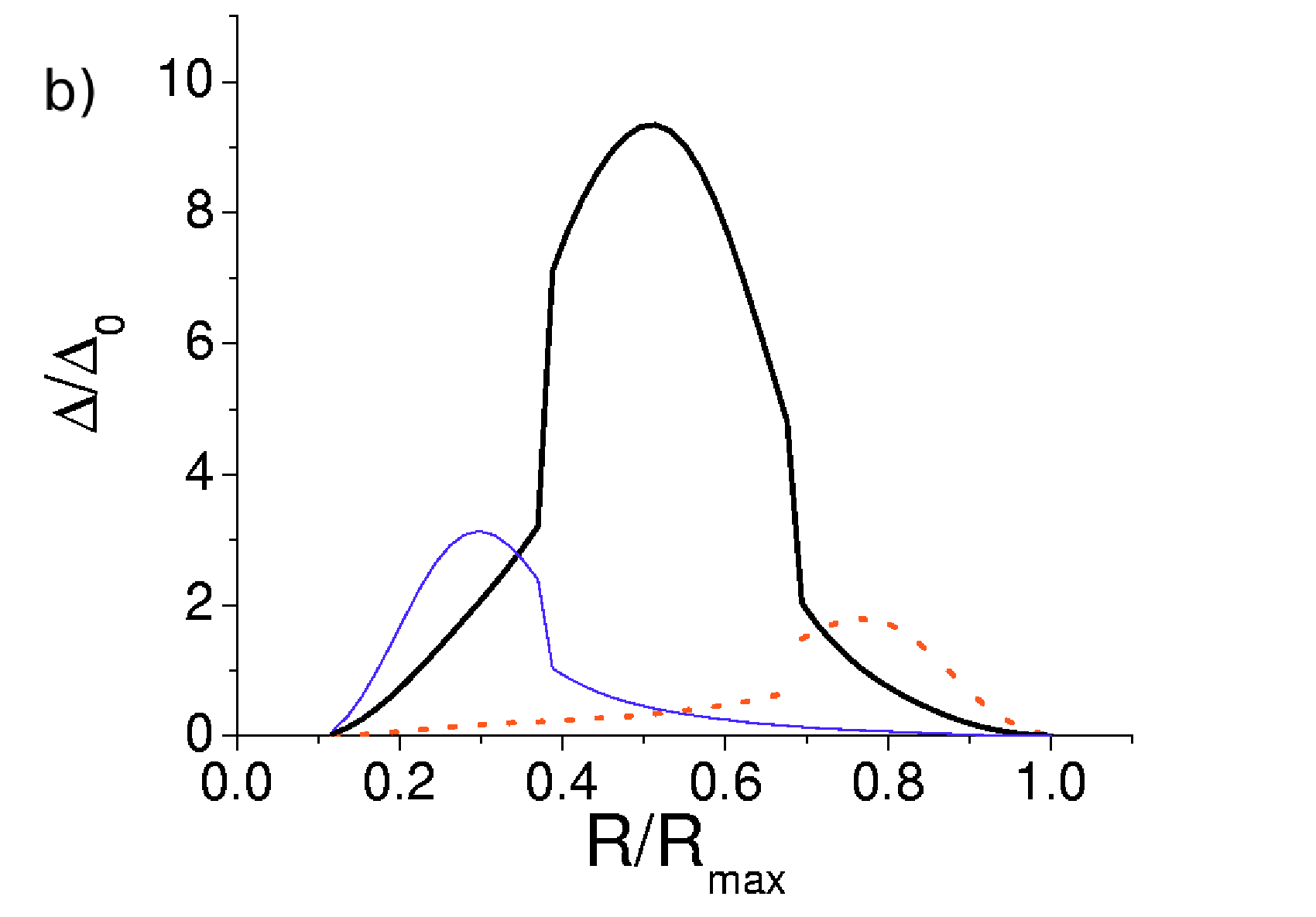}
\caption{\label{fig4} \footnotesize{The geometry of the system is
depicted in Fig. 3a and consists of $3$ axially symmetric layers.
The order parameter $\Delta(\rho)$ (see Fig. 4b) for a nonuniform
hollow wire with three layers and three different arrangements,
the better superconductor has a doubled coupling constant.
$R_{min}=1.5 nm$ and $R_{max}=15$ nm. The length of the wire is
chosen to be $500$ nm. For our calculations we assume temperature
$T=0$ K. }}
\end{figure}

Now let us consider an inhomogeneous wire composed of three
axially symmetric layers. In Fig. \ref{fig4} we compared three
different arrangements for the three layer system. We have chosen
the layers to be of the same thickness, i.e. $g(\rho)=g_1$, for
$R_{min}<R<(R_{max}+2 R_{min})/3$, $g(\rho)=g_2$, for $(R_{max}+2
R_{min})/3<R<(2 R_{max}+R_{min})/3$ and $g(\rho)=g_3$, for $(2
R_{max}+R_{min})/3<R<R_{max}$. The coupling constant for the one
of the layers (either $g_1$ or $g_2$ or $g_3$) is set to $2g$, and
the two others are set to $g$. For this type of inhomogeneity the
best choice that maximizes the superconducting order parameter
$\Delta(\rho)$ is the one, where the better superconductor is in
the middle. Since the layers considered here are thinner compared
to the layers shown in Fig. \ref{fig3}, the superconducting order
parameter of the layer closest to the inner surface is
significantly reduced by the proximity effect of the nearest
boundary with $\Delta(R_{min})=0$.

Now let us consider an optimal design problem. It is known, that
the condensation energy $E_{cond}$ is an important  quantity that
characterizes superconducting systems. Obviously, for
inhomogeneous system it may vary for different arrangement of the
superconducting layers. It is interesting to find an optimal
configuration that maximizes $E_{cond}$.

In order to isolate volume effects we assume a superconducting
layer of a fixed volume, that is equal to one third of the total
volume of the wire: $V_{2g}=\pi(R^2_{max}-R^2_{min})L/3$. This
layer has the coupling constant $2g$. The rest material in the
wire has the coupling constant $g$. The problem is to find the
best placement of the layer to maximize the condensation energy
$E_{cond}$ \bea \label{cond} E_{cond} \propto
\int_{R_{min}}^{R_{max}} \Delta^2(\rho) \rho d\rho. \eea

Let the position of the layer's boundary, that is closest to the
axis of the symmetry labelled by $R$, then the layer's boundary
closest to the outer surface of the wire will be
$R_r=\sqrt{V_{2g}/(\pi L)-R^2}$ where $L$ is the length of the
wire.

\begin{figure}[h] %FIGURE 2
%\vspace{1.cm}
\includegraphics[width=10.cm, angle=-90]{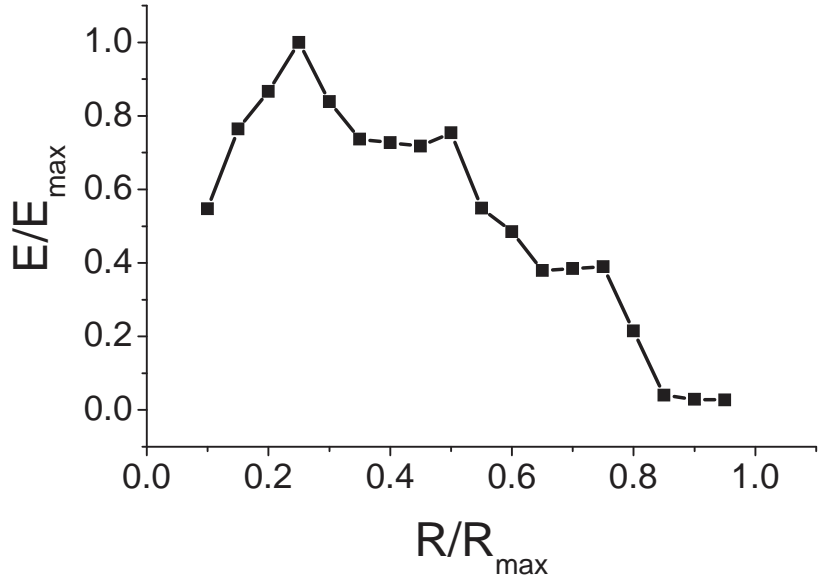}
\caption{\label{fig7} \footnotesize{The geometry of the system
under consideration is similar to what is shown in Fig 5a. The
better superconducting layer is in the middle. The condensation
energy $E \propto \int \Delta^2(\rho)\rho d\rho$ of a
heterostructure as a function of the position of the better
superconducting layer. The better superconductor has a doubled
coupling constant $2g$, the worse superconducting layers have $g$.
$R_{min}=1.5 nm$ and $R_{max}=15$ nm. The length of the wire is
chosen to be $500$ nm. For our calculations we assume temperature
$T=0$ K. }}
\end{figure}

In Fig. \ref{fig7} we plot the condensation energy $E_{cond}$
normalized to its maximum value $E_{max}$, as a function of its
position $R$. For our simulations we choose $R_{min}=1.5$ nm,
$R_{max}=15$ nm, $L=500$ nm. One can clearly observe that the
condensation energy has a maximum near $R\approx 5$ nm. For this
placement the reducing effect from the boundary
$\Delta(R_{min})=0$ is minimized, while the effective enhancement
of the superconducting order parameter due to the local curvature
is still taking place. There are some other local maxima of the
condensation energy, that we attribute to finite size resonance
\cite{cyl1}.

\begin{figure}[h] %FIGURE 2
%\vspace{1.cm}
\includegraphics[width=7.cm, angle=-90]{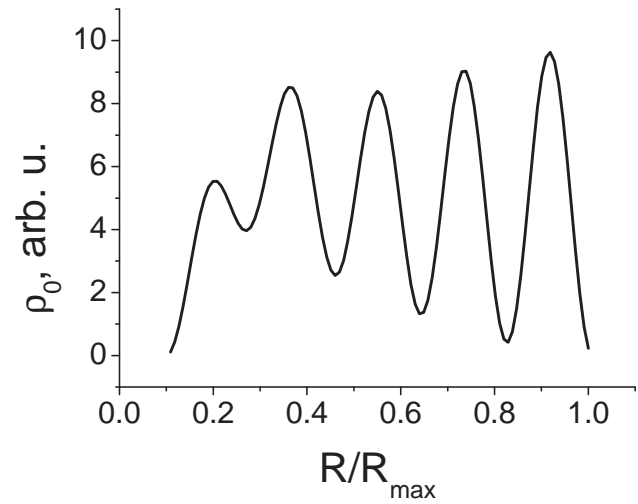}
\caption{\label{fig5} \footnotesize{Local density of states (LDOS)
$\rho_0(\rho,E)$, for a uniform hollow wire with $R_{min}=1.5$ nm
and $R_{max}=15$ nm, $E=\Delta_{bulk}$. The length of the wire is
chosen to be $500$ nm. For our calculations we assume temperature
$T=0$ K. }}
\end{figure}

As we mentioned above, it is now possible to measure the local
density of states (LDOS) directly using STM technique. In the STM
experiment \cite{STM} LDOS is proportional to the local
differential tunneling conductance. In Fig. \ref{fig5} we plot the
calculated local density of states $\rho_0$ for a homogeneous
superconducting nanowire that is determined through \bea
\label{ldos}
\rho_0(\rho,E)=\sum_n[|u_n(\rho)|^2\delta(E-E_n)+|v_n(\rho)|^2\delta(E+E_n)].
\eea We again consider a hollow wire with $R_{min}=1.5$ nm and
$R_{max}=15$ nm. The length of the wire is chosen to be $500$ nm.
We use $E=\Delta_{bulk}$ for these calculations. In Fig.
\ref{fig5} one can see relatively big spatial variations of the
local density of states due to the confinement of quasiparticles
in the cylindrically symmetric boundaries. One possibly may
observe these variations by scanning the end face of the wire by a
STM tip.

%In Fig. \ref{fig6} we also plot LDOS $N_0$ as a function of
%$\rho$ and scanned energy $E$ (preliminary results).

\section{Discussion}

The notion of design of quantum properties is well established.
The examples of this quantum state optimization include band
structure engineering \cite{zunder1,zunder2}, engineering of
complex materials structure \cite{ceder}. Notion of optimal design
also can be viewed as similar to the ideas of
decoherence-protected quantum state engineering in quantum
computing \cite{eberly,ilya}.

We proposed to apply these ideas to the superconductors as one
example of macroscopic quantum state. The optimal design approach
allows us to investigate the optimal structures that maximize the
functional property of nanowires, like maximal condensate energy
and superconducting gap.

We find that depending on the desired property, specific design is
different. Specifically we find that for two layered
superconducting wires  maximal gap is obtained when the
superconductor with the stronger pairing potential is placed on
the inside. The reason for this enhancement is a confinement
effect in the cylindrical geometry.

In the case one wants to optimize the maximum critical current one
would need to place stronger SC on the outside because this will
be the region where the screening current flows the most. We
explicitly demonstrated that depending on the function one wants
to optimize the superconducting wire design {\em with the same
ingredients} can be very different. On the other hand, in the case
of three layer superconductors we see that the largest gap (with
enhancement factor of about 300\%) is obtained when the strongest
superconductor is placed in the middle of a sandwich.

Optimal design ideas presented here are applicable to other
configurations. One might look at the multilayered structures
obtained by the molecular beam epitaxy and other methods. Here the
combination of superconducting and insulating or magnetic layers
(ferromagnetic, multiferroic and other layers) would allow us to
investigate the optimal design of multilayered structures with
competing interactions. One particularly interesting case is the
experiment by Bosovic {\it et al} \cite{BosovicNature}. In their
multilayered configuration  pseudogaped normal regions are
alternating with the superconducting regions in multilayers. The
unusual proximity effect at very large distances of up to 100 \AA
 has been observed. We can take these configurations as a guidance
on how possible future multilayered structures  might look. The
ideas of optimizing the competing interactions in these
configurations are becoming relevant.

\section{Conclusion}
We consider spatially inhomogeneous superconducting materials of
the axial symmetry. We also studied the interplay between local
curvature and spatial inhomogeneity of the system. We find that
due to the cylindrical geometry, different arrangement of
superconducting layers are not equivalent because of the different
local curvature, and the interchange of the sequence of the layers
gives different results. We find that placing of the better
superconducting material layer closer to the symmetry axis results
in the higher values of the pairing potential. However, it is not
true in the case when this layer is thin enough. In this case the
boundary effects will reduce effect of the small local curvature.
We have proved this for the case of three layers.

We also studied the optimal design problem of placing of the
better superconducting layer to achieve maximum pairing potential.
The optimal placement is a result of compromise between the local
curvature effects and effect of the close boundaries. For more
complicated geometries and larger number of layers one should
perform search with the help of global optimization techniques,
like Genetic Algorithms. The obtained solution are most likely
will not be accessible with any known analytical methods.

For ultrathin nanostructures we considered in this work, localized
impurity effects and phase fluctuations may play a significant role.
The impurities and defects can be modeled on the mean field level by
introducing localized impurity potentials, similar how it was done,
for example, in \cite{sasha_prl}. The phase fluctuations effects may
compete with the pairing potential enhancement due to quantum
confinement of quasiparticles in the cylindrical geometry of a
nanowire. However, such effect goes beyond mean field BdG theory
used in this work. The effect of quantum fluctuations may be an
interesting topic to study in the future.

As a possible way to extend our research, we consider to study
nonequilibrium  transport properties of inhomogeneous nanowires
(super-current distribution, etc.). Since the nanowires under our
considerations are much smaller than the characteristic field
screening length, the inhomogeneity of the pairing potential will
not play a significant role for relatively small fields and
currents. However, for larger currents, closer to the critical
value, the geometry and size effects may significantly change
nonequilibrium characteristics of inhomogeneous superconducting
nanowires, resulting in higher values of the critical current. It
may also be interesting to study how the critical current value
depends on the length of nanowire, and how the pairing potential
changes along the nanowire of a finite length.

We believe, the optimal design has a bright future in nanoscience
since it can potentially convert the available computational power
into improvement of target properties of rationally structured
materials and may substantially increase efficiency of engineered
devices.
\section{Acknowledgements}

This work was carried out under the auspices of the National
Nuclear Security Administration of the U.S. Department of Energy
at Los Alamos National Laboratory under Contract No.
DE-AC52-06NA25396. \setcounter{equation}{0}
\renewcommand{\theequation}{A-\arabic{equation}}
\section*{APPENDIX}

We already outlined in the introduction the main points of our
approach. Here we will elaborate on a more technical details of
our numerical approach. In our simulations the radial part of the
Bogoliubov-de Gennes equations Eq.(\ref{BDG_cyl}) has a
singularity at the origin $\rho=0$ that needs uneven discreization
mesh and mixed type of the boundary conditions \cite{cyl1}. In
order to avoid these complications, we restricted ourselves to
consider hollow cylinders with inner radius $R_{min}$ and outer
radius $R_{max}$. We assume Dirichlet type of the boundary
conditions, namely: \bea \label{Dirichlet}
u_n(R_{min})=u_n(R_{max})=v_n(R_{min})=v_n(R_{max})=0. \eea

Then we introduce a transformation \bea \label{transformation}
\hat{a}(\rho)=\hat{b}(\rho)/\sqrt{\rho} \eea that removes in
Eq.(\ref{BDG_cyl}) the term with the first derivative with respect
to $\rho$: \bea \label{BDG_symm} ( -\frac{\hbar^2}{2
m^*}\partial^2/\partial \rho^2 -(l^2-1/4)\rho^{-2}-k_z^2-\mu
)\hat{\sigma}_z \hat{b}_n(\rho)+\Delta(\rho)\hat{\sigma}_x
\hat{b}_n(\rho)=E_n \hat{b}_n(\rho). \eea The transformation
Eq.(\ref{transformation}) guarantees that the descretized
Hamiltonian in Eq.(\ref{BDG_symm}) is represented by a symmetric
matrix, that ensures much faster numerical diagonalization.

The radial part of the Bogoliubov-de Gennes equations
Eq.(\ref{BDG_symm}) is discretized on a equidistant mesh in real
space. For this purpose we use the $13$th order 12-point
discretization rule for the second derivative \cite{scheme}. The
exact diagonalization of the resulting sparse matrix is then
performed numerically using the ARPACK eigensolver \cite{arpack}
available freely on the web \cite{web}. We performed a comparison
of the numerical results and the analytical solution available for
$\Delta(\rho)\equiv0$, and find a good agreement using the number
of discretization points $N=200$.

For every iteration step of the self-consistent calculations, the
numerical diagonalization procedure is performed for different
values of the wave vector $k_z$ and the azimuthal quantum number
$l$. The procedure stops when no eigenvalues lay in the Debye
window determined by $\hbar \omega_D$.


%\begin{figure}[h] %FIGURE 2
% \vspace{1.cm}
%\includegraphics[width=7.cm, angle=-90]{density.ps}
%\caption{\label{fig3} \footnotesize{Density of states for
%cylindrical symmetry. $R_{min}=1.5$ nm, $R_{max}=15$ nm.
% }}
%\end{figure}

% \begin{figure}[h] %FIGURE 2
% \vspace{1.cm}
%\includegraphics[width=7.cm, angle=-90]{3D.ps}
%\caption{\label{fig6} \footnotesize{Local density of states (LDOS)
%$N_0(\rho,E)$ for a uniform hollow wire with $R_{min}=1.5 nm$ and
%$R_{max}=15$ nm. For our calculations we assume temperature $T=0$
%K.
% }}
%\end{figure}

\end{document}